\documentclass[preprint,preprintnumbers,amsmath,amssymb,floatfix,nofootinbib,h-physrev.bst]{revtex4}
\usepackage{epsf, array, color}
\usepackage{graphicx}
\usepackage{subfigure}
\usepackage{bbold}
\usepackage{amsmath}
\usepackage{mathtools}
\usepackage{dcolumn}

\newcommand{\bea}{\begin{eqnarray}}
\newcommand{\eea}{\end{eqnarray}}

\def\beq{\begin{equation}}
\def\eeq{\end{equation}}

\newcommand{\lsim}{\lesssim}

\begin{document}

\title{NLO Corrections to Double Higgs Production in  the Higgs Singlet Model}

\author{S.~Dawson$^{\, a}$ and I.~M.~Lewis$^{\, b}$ }
\affiliation{
\vspace*{.5cm}
  \mbox{$^a$Department of Physics,\\
  Brookhaven National Laboratory, Upton, N.Y., 11973,  U.S.A.}\\
 \mbox{$^b$ SLAC National Accelerator Laboratory,\\
 2575 Sand Hill Rd, Menlo Park, CA, 94025, U.S.A.}
 \vspace*{1cm}}

\date{\today}

\begin{abstract}
Higgs pair production at the LHC from gluon fusion is small in the Standard Model, but can be 
 enhanced in models where a resonant enhancement is allowed.  We examine the effect
of a resonant contribution from a second scalar arising in a model with a gauge singlet scalar field
in addition to the usual $SU(2)$ scalar doublet, with mass up to $M_H\sim 600~GeV$ and
discuss the interference effects in double Higgs production.  The interference effects distort the double Higgs invariant mass distributions, and, depending on $M_H$, can enhance the total cross section by up to $\sim20\%$ or decrease by $\sim 30\%$ for viable mixing parameters.  We compute the NLO QCD corrections in the large $m_t$
limit.  The corrections are large and can also significantly distort kinematic distributions near the resonance peak.
\end{abstract}

\maketitle

\section{Introduction}
The experimental exploration of the Higgs sector of the Standard Model (SM) is one of the main goals of the current LHC run.  Current data
on Higgs properties are in reasonable agreement with the theoretical expectations, although there is still considerable room for new physics.  
An attractive extension of the SM is the Higgs portal scenario, in which the SM Higgs boson couples  to a gauge singlet scalar, $S$, which in turn can 
communicate with a hidden sector.  Models with an additional scalar singlet have also been used to generate
a strong first order electroweak phase transition\cite{Profumo:2014opa,Curtin:2014jma,Espinosa:2011ax,No:2013wsa,Profumo:2007wc,Kozaczuk:2015owa}.

In the Higgs singlet model, the SM Higgs doublet mixes with the new singlet, $S$, to form
two physical scalar bosons: one, $h$, identified with the observed $m_h=125 ~GeV$ resonance and a second, $H$, with mass $M_H$.  When $M_H\gtrsim 2 m_h$, large resonant enhancements are possible in double Higgs production 
from gluon fusion, significantly enhancing the rate compared to the SM prediction.  The singlet model has the advantage of depending
on relatively few parameters, allowing for straightforward experimental study at the LHC in the 
analysis of Higgs couplings\cite{ATLAS-CONF-2014-010}, 
searches for heavy SM-like Higgs bosons\cite{ATLAS:2014aga,ATLAS-CONF-2013-030,Khachatryan:2015cwa} and direct searches
for resonant di-Higgs production\cite{Aad:2015uka,Aad:2014yja,CMS-PAS-HIG-13-032,Khachatryan:2015yea}.
Higgs singlet models have also been extensively studied theoretically
 and additional limits derived from precision electroweak data, the interpretation of 
 LHC results, and restrictions from the requirements of perturbative unitarity and perturbativity of the couplings\cite{Robens:2015gla,Pruna:2013bma,Bowen:2007ia,Espinosa:2011ax,Dolan:2012ac,Chen:2014ask,Falkowski:2015iwa,O'Connell:2006wi,Barger:2007im,Englert:2013gz,Englert:2014ffa,
Englert:2014uua,Dolan:2015zja,Englert:2014uqa,Buttazzo:2015bka,Martin-Lozano:2015dja,Godunov:2015nea}. 
 
Double Higgs production from gluon fusion in the SM results from both 
triangle and box loop contributions, which interfere destructively, causing a
suppression of the total rate from the naive estimate\cite{Plehn:1996wb,Glover:1987nx}. 
This process 
 has been
studied at lowest order QCD (LO) in the singlet model, and regions of parameter space with enhanced rates determined. In this work, we consider precision predictions at NLO QCD
 for double Higgs production in the singlet model, including the $hh$ invariant mass distribution.
 Since double Higgs production from gluon fusion
first occurs at one-loop, the full NLO corrections involve two-loop virtual diagrams with massive internal particles.  The calculation is
considerably simplified by using an effective theory corresponding to the $m_t\rightarrow \infty$ limit of the SM.  In the SM, the corrections to the total rate have been
known at NLO for some time in the effective theory\cite{Dawson:1998py}, which has also been matched onto the NNLL threshold resummed result~\cite{Shao:2013bz}.  Recently the rate has been calculated at NNLO\cite{deFlorian:2013jea,Grigo:2014jma} and matched to the 
 NNLL result\cite{deFlorian:2015moa}.
  These corrections typically
increase the rate by a factor of about $2-2.3$. The SM NLO QCD corrections 
to $gg\rightarrow hh$ are also  known in an effective field
theory limit where the exact mass dependence is retained everywhere except in the virtual 
corrections\cite{Frederix:2014hta} and alternatively in an expansion in 
${1\over m_t^{2n}}$\cite{Grigo:2013xya,Grigo:2013rya}. 
The unknown $m_t$ dependence of the higher order QCD corrections induces an uncertainty of ${\cal O}(\pm 10\%)$ in the SM predictions.

Higher order QCD corrections to new physics scenarios with resonant enhancements of the double
Higgs rates have been derived for the MSSM\cite{Baglio:2012np,Dawson:1998py}
 and the two Higgs doublet model\cite{Hespel:2014sla}, and also in an effective operator
formalism with no resonance\cite{Grober:2015cwa}.  
These corrections not only affect the total rate, but in some regions of parameter space distort the shape of the distributions.  In this paper, we examine
the approximations behind the QCD corrections in the context of the Higgs singlet model.  We demonstrate that the corrections in the resonance region
are significant and that the use of a constant $K$ factor is a poor approximation in this regime. We also investigate the interference effects between the heavy scalar and SM-like contributions.  These effects can be significant and should be included in searches for new heavy scalars.

\section{Model}
\label{sec:model}
\subsection{Recap}
We consider a simple extension of the SM  containing the SM Higgs doublet, $\Phi$, and an additional
real gauge singlet scalar,  $S$.  After imposing a $Z_2$ symmetry under which $S\rightarrow -S$, 
the most general scalar potential is\cite{Robens:2015gla,Bowen:2007ia}
\begin{eqnarray}
V &=& - \mu^2 \, \Phi^\dagger \Phi -m^2 S^2 +\lambda (\Phi^\dagger \Phi)^2
   + \frac{a_2}{2} \, \Phi^\dagger \Phi \, S^2   + \frac{b_4}{4} S^4.
\label{potential}
\end{eqnarray}
Although not necessary for a strong first order electroweak phase transition, models without a $Z_2$ symmetry have been constructed in the context of electroweak 
baryogenesis\cite{Profumo:2014opa,Curtin:2014jma,Espinosa:2011ax,No:2013wsa,Profumo:2007wc,Kozaczuk:2015owa}.  However, the additional complication is not 
necessary for our discussion of higher order corrections. 
After spontaneous symmetry breaking,  in the unitary gauge we have $\Phi^{\rm T}=(0,\phi_0)/\sqrt{2}$ with $\langle \phi_0\rangle \equiv v=246~GeV$
and $S\equiv(s+x)/\sqrt{2}$ with $\langle S\rangle=x/\sqrt{2}$.

The mass eigenstate fields, $h$ and $H$, are:
\begin{equation}
\left(
\begin{array}{c}
h \\ H\end{array}
\right)=\left(\begin{array}{cc} 
\cos\theta & -\sin\theta \\
\sin\theta & \cos \theta
\end{array}
\right)
\left( 
\begin{array}{c}
\phi_0-v\\
s
\end{array}
\right)\, ,
\end{equation}
with physical masses, $m_h$ and $M_H$, and $-{\pi\over 2}\le \theta \le {\pi\over 2}$.

The terms in the potential can be written in terms of the physical masses and mixing angle as,
\begin{eqnarray}
\mu^2 &=& v^2 \lambda+\frac{1}{4}x^2 a_2\\
m^2 &=&\frac{1}{4}\left(x^2 b_4+v^2 a_2\right)\\
\lambda&=&{m_h^2\over 2 v^2}  +{M_H^2-m_h^2\over 2 v^2}\sin^2\theta\\
a_2&=& {M_H^2-m_h^2\over v x}(2\sin\theta\cos\theta) \\
b_4&=& {2M_H^2\over x^2}+{2(m_h^2-M_H^2)\over x^2}\sin^2\theta\, .
\end{eqnarray}
The requirement that the potential be bounded from below imposes,
\begin{equation}
a_2 >  -2\sqrt{b_4 \lambda}\, , \qquad \lambda, b_4 > 0\, .
\end{equation}
We will also need the triple scalar couplings:
\begin{equation}
L\sim {\lambda_{111}\over 6} h^3+{\lambda_{211}\over 2}H h^2+...
\end{equation}
where
\begin{eqnarray}
\lambda_{111}&=& -{3m_h^2\over v}\biggl(\cos^3\theta-\tan\beta\sin^3\theta\biggr)\label{lam111}\\
\lambda_{211}&=&-{m_h^2\over v}\sin 2 \theta(\cos\theta+\sin\theta\tan\beta)\biggl(1+{M_H^2\over 2 m_h^2}\biggr)\label{lam211}
\end{eqnarray}
and  $\tan\beta\equiv {v\over x}$. A complete list of the scalar self-couplings can be found 
in the Appendix of Ref. \cite{Chen:2014ask}.

\begin{figure}
\begin{centering}
\includegraphics[scale=0.6]{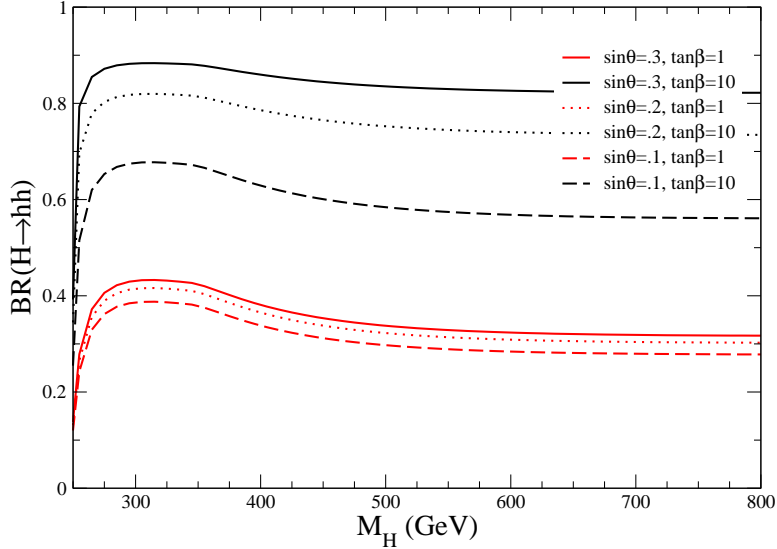}
\par\end{centering}
\caption{Branching ratio for $H\rightarrow hh$.}
\label{fig:Hbr}
\end{figure}

We assume that the lightest   scalar, $h$, is the SM-like Higgs particle with $m_h=125~GeV$.  The decay widths  to SM
particles, $X$, are then simply the SM values rescaled by the scalar mixing angle,
\begin{eqnarray} 
\Gamma(h\rightarrow XX^\dagger)&=&\cos^2\theta \Gamma(h\rightarrow XX^\dagger)_{SM}\nonumber \\
\Gamma(H\rightarrow XX^\dagger)&=&\sin^2\theta\Gamma(H\rightarrow XX^\dagger)_{SM}
\end{eqnarray}
where $\Gamma(H\rightarrow XX^\dagger)_{SM}$ is the SM partial width evaluated at mass $M_H$.  The total widths are
\begin{eqnarray}
\Gamma_h &=& \cos^2\theta \Gamma^{SM}_h\nonumber\\
\Gamma_H &=& \sin^2\theta \Gamma^{SM}_H+\theta(M_H-2 m_H)\Gamma(H\rightarrow hh),
\end{eqnarray}
where $\Gamma^{SM}_H$ is the SM total width evaluated at mass $M_H$ 
and
\begin{equation}
\Gamma(H\rightarrow hh)={\lambda_{211}^2\over 32\pi M_H}\sqrt{1-{4m_h^2\over M_H^2}}\, .
\end{equation}
The branching ratio of $H\rightarrow hh$ is shown in Fig. \ref{fig:Hbr}. For small $\sin\theta$, the branching ratio is relatively insensitive to $\tan\beta$ and is approximately ${\rm BR}(H\rightarrow hh)\sim 0.3-0.4$.

The model has $5$ free parameters which we take to be:
\begin{equation}
m_h=125~GeV, M_H, v=246~GeV, \tan\beta, \cos\theta\, .
\end{equation}

\subsection{Limits}
The $Z_2$ symmetric Higgs singlet model is restricted by a number of experimental measurements.  Fits
to the $h$  couplings assuming no branching ratio to invisible particles require $\mid \cos\theta\mid  > 0.93$ at $95\%$ confidence
level\cite{ATLAS-CONF-2014-010}.  Precision electroweak quantities\cite{Dawson:2009yx}, in particular the $W$ boson mass\cite{Lopez-Val:2014jva}, receive contributions which are sensitive
to $M_H$ and $\cos\theta$.  For $M_H\gtrsim 400~ GeV$, measurements of the $W$ mass require $\mid \cos\theta \mid > 0.96$, with the
limits significantly weaker for smaller $M_H$\cite{Robens:2015gla,Falkowski:2015iwa}.   Heavy Higgs searches can also be interpreted as limits on $\cos\theta$.  For 
$ M_H \lesssim 300~GeV$, these limits are stronger than the limits from the $W$ mass.  Assuming no branching ratio, $H\rightarrow hh$,
the direct search limits for heavy Higgs bosons can be interpreted as requiring $\mid \cos\theta\mid > 0.92$ in this region.  Requiring $b_4$ to remain perturbative
as it is scaled to high energy gives an upper limit on $\tan\beta$ which depends on $M_H$ and $\theta$:  for $\sin\theta=0.1$ and $M_H =200 (500)~GeV$,~$\tan\beta <1.5 (0.5)$\cite{Pruna:2013bma,Falkowski:2015iwa}.
With these considerations in mind, we will in general present results with $\cos\theta=0.96, \tan \beta=0.5$.

\section{Double Higgs Production}
\subsection{LO Results}
\begin{figure}
\begin{centering}
\includegraphics[scale=0.4]{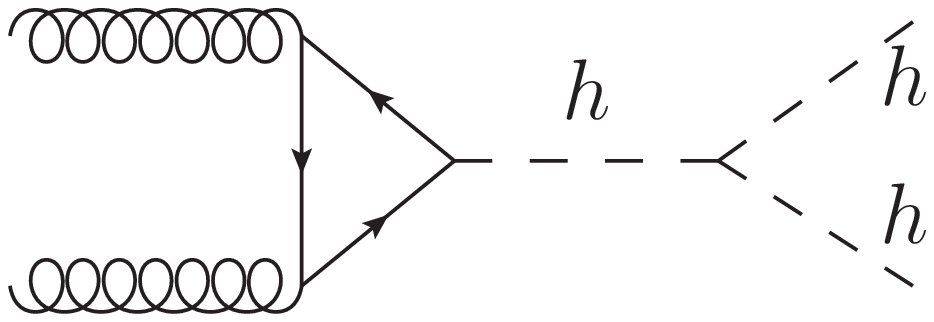}
\includegraphics[scale=0.4]{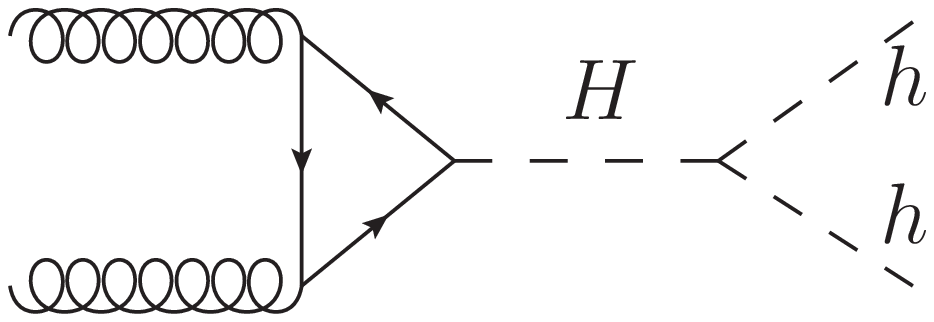}
\includegraphics[scale=0.4]{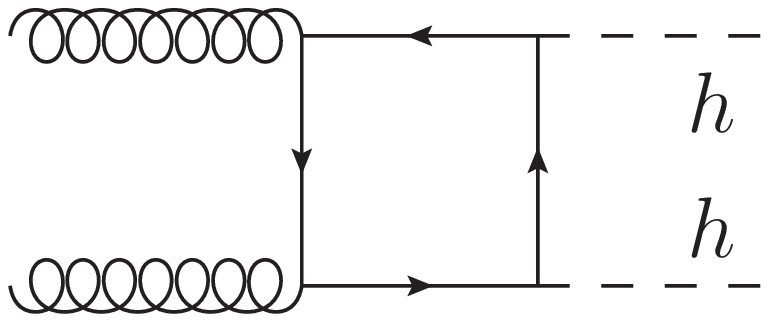}
\par\end{centering}
\caption{Feynman diagrams contributing to $gg\rightarrow hh$ in 
the singlet model.}
\label{fig:feyndiags}
\end{figure}
Two Higgs production arises from the Feynman diagrams shown in Fig. \ref{fig:feyndiags}.
The result is sensitive to new colored objects  with mass $m$ (fermions or scalars) in the loops
\cite{Dawson:2015oha,Kribs:2012kz,Asakawa:2010xj,Chen:2014xwa,Dawson:2012mk,Batell:2015koa}
 and also to the
$3-$ Higgs self-couplings.
The amplitude for $g^{A,\mu}(p)g^{B,\nu}(q)\rightarrow h(p^\prime)h(q^\prime)$ can be
written as,
\begin{equation}
A^{\mu\nu}_{AB}={\alpha_s\over 8 \pi v^2}\delta_{AB}\biggl(P_1^{\mu\nu}(p,q)F_1(s,t,u,m^2)+
P_2^{\mu\nu}(p,q,p^\prime)F_2(s,t,u,m^2)\biggr)\, ,
\end{equation}
where $P_1$ and $P_2$ are the orthogonal  projections onto the spin-$0$ and spin-$2$  states respectively,
\begin{eqnarray}
P_1^{\mu\nu}(p,q)&=&
g^{\mu\nu}-{p^\nu q^\mu\over p\cdot q}\nonumber \\
P_2^{\mu\nu}(p,q,p^\prime)&=&g^{\mu\nu}+{2\over s p_T^2 }
\biggl\{
m_h^2 p^\nu q^\mu
+(t-m_h^2) q^\mu p^{\prime \nu}
+(u-m_h^2) p^\nu p^{\prime \mu}
+ s p^{\prime \mu} p^{\prime \nu}
\biggr\}\, ,
\end{eqnarray}
$s,t$, and $u$ are the  partonic Mandelstam variables,
\begin{eqnarray}
s&=&(p+q)^2\nonumber \\
t&=&(p-p^\prime)^2\nonumber \\
u&=&(p-q^\prime)^2
\nonumber \\
p_T^2&=&\biggl({ut-m_h^4\over s}\biggr)\, .
\end{eqnarray}
The functions $F_1$ and $F_2$ are known analytically\cite{Plehn:1996wb,Glover:1987nx}
and
the partonic cross section is given in terms of the form factors by,
\begin{eqnarray}
{d{\hat\sigma}_{LO}^{m_t}\over dt}&=&{\alpha_s^2(\mu_R)\over 2^{15}\pi^3v^4}
\biggl(
{\mid F_1(s,t,u,m_t^2)\mid^2+\mid F_2(s,t,u,m_t^2) \mid^2\over s^2}\biggr)\,,
\end{eqnarray}
where $\mu_R$ is the renormalization scale.  
(We have included the factor of ${1\over 2}$ for identical particles in the final
state).
In the singlet model (as in the SM), the dominant contribution comes from top quark loops.
The form factors can be written as,
\begin{eqnarray}
F_1(s,t,u,m_t^2)&\equiv &F_1^{tri}(s,t,u,m_t^2)+F_1^{box}(s,t,u,m_t^2)\nonumber \\
F_1^{tri}(s,t,u,m_t^2)&=&-
s\biggl({\cos\theta \lambda_{111}v\over s-m_h^2+im_h\Gamma_h}
+{\sin\theta \lambda_{211}v\over s-M_H^2+iM_H\Gamma_H}\biggr)
F_{\triangle}(s,m_t^2)
\nonumber \\
F_1^{box} (s,t,u,m_t^2)&=&s \cos^2\theta
F_{\square}(s,t,u,m_t^2)
\nonumber \\
F_2(s,t,u,m_t^2)&=&s\cos^2\theta
G_{\square}(s,t,u,m_t^2)\, .
\label{formdef}
\end{eqnarray}
In the limit $m_t\rightarrow \infty$,
\begin{eqnarray}
F_\triangle&\rightarrow &{4\over 3}\nonumber \\
F_\square &\rightarrow &-{4\over 3} \nonumber \\
G_\square&\rightarrow &0\,.
\label{formlims}
\end{eqnarray}
The form factors $F_\triangle,F_\square$, and $G_\square$  including the full kinematic dependences are found in Refs. \cite{Plehn:1996wb,Glover:1987nx}\footnote{ The functions defined in Eq. \ref{formdef} satisfy 
$F_\triangle (F_\square,G_\square) \rightarrow 2 F_\triangle (F_\square, G_\square)$(Ref. \cite{ Plehn:1996wb}) and $s F_\triangle (sF_\square,sG_\square) \rightarrow  F_\triangle (F_\square, G_\square)$(Ref. \cite{Glover:1987nx}).}.
We denote the cross section found by including the exact $m_t$ dependence of the matrix
elements, Eq. \ref{formdef},
by ${\hat{\sigma}}_{LO}^{m_t}$, 
and the $m_t\rightarrow\infty$ limit, Eq. \ref{formlims}, as ${\hat{\sigma}}_{LO}^{m_t\rightarrow
\infty}$.

The LO hadronic cross section is,
\begin{equation}
\sigma_{\mathrm{LO}}^{m_t}  =  \int_{\tau_0}^1 d\tau~\frac{d{\cal L}^{gg}}{d\tau}~
\hat\sigma_{\mathrm{LO}}^{m_t}(s = \tau S)\, ,
\label{sigmlo}
\end{equation}
and the luminosity function is defined,
\begin{equation}
{d{\cal L}^{ij}
\over d\tau}=\sum_{ij} \int_{\tau}^1{dx\over x} f_i(x,\mu_F) f_j\biggl({\tau\over x},\mu_F\biggr)\, ,
\end{equation}
$S$ is the square of the hadronic energy, $\tau_0 = \frac{4m_h^2}{S}$,
 and $\mu_F$ is
the factorization scale.

\subsection{NLO Corrections}
\label{sec:NLO}

The NLO corrections in the SM are known in the large $m_t$ limit\cite{Dawson:1998py} and
are trivially generalized to the singlet model. The $gg$ initial state contains IR singularities which cancel when the real and virtual 
contributions are combined.  The remaining collinear divergences in the $gg$, $qg$ and $q {\overline q}$ initial states are absorbed into the NLO PDFs defined
in the ${\overline {MS}}$ scheme with $5$ light flavors.  The terms listed below
are the finite contributions obtained after canceling the singularities. We write the NLO rate as,
\begin{equation}
\sigma_{\mathrm{NLO}}^{m_t}(pp \rightarrow hh) = 
\sigma_{\mathrm{LO}}^{m_t} + 
\sigma_{\mathrm{virt}}^{m_t} + \sigma_{gg}^{m_t}
 + \sigma_{gq}^{m_t} + \sigma_{q\bar{q}}^{m_t},
\label{eq:gghqcd}
\end{equation}
where,
\begin{eqnarray}
 \sigma_{\mathrm{virt}}^{m_t} & = & \frac{\alpha_s(\mu_R)} {\pi}\int_{\tau_0}^1 d\tau~
\frac{d{\cal L}^{gg}}{d\tau}~\hat \sigma_{\mathrm{LO}}^{m_t}(s=\tau S)~C^{m_t}, 
\nonumber \\ 
\sigma_{gg}^{m_t} & = & \frac{\alpha_{s}(\mu_R)} {\pi} \int_{\tau_0}^1 d\tau~
\frac{d{\cal L}^{gg}}{d\tau} \int_{\tau_0/\tau}^1 \frac{dz}{z}~
\hat\sigma_{\mathrm{LO}}^{m_t}(s = z \tau S)
\left\{ - z P_{gg} (z) \log \frac{\mu_F^{2}}{\tau s} \right. 
\nonumber \\
& & \left. \hspace*{3.0cm} {} - \frac{11}{2} (1-z)^3 + 6 [1+z^4+(1-z)^4]
\left(\frac{\log (1-z)}{1-z} \right)_+ \right\}, 
\nonumber \\ 
 \sigma_{gq}^{m_t} & = & \frac{\alpha_{s}(\mu_R)} {\pi} \int_{\tau_0}^1 d\tau
\sum_{q,\bar{q}} \frac{d{\cal L}^{gq}}{d\tau} \int_{\tau_0/\tau}^1 \frac{dz}{z}~
\hat \sigma_{\mathrm{LO}}^{m_t}(s = z \tau S)
\left\{ -\frac{z}{2} P_{gq}(z) \log\frac{\mu_F^{2}}{\tau s(1-z)^2} \right. 
\nonumber \\
& & \left. \hspace*{9.0cm} {} + \frac{2}{3}z^2 - (1-z)^2 
\vphantom{\frac{M^{2}}{\tau s(1-z)^2}} \right\},
\nonumber \\ 
 \sigma_{q\bar q}^{m_t} & = & \frac{\alpha_s(\mu_R)}
{\pi} \int_{\tau_0}^1 d\tau
\sum_{q} \frac{d{\cal L}^{q\bar q}}{d\tau} \int_{\tau_0/\tau}^1 \frac{dz}{z}~
\hat \sigma_{\mathrm{LO}}^{m_t}(s = z \tau S)~\frac{32}{27} (1-z)^3\, .
\label{eq:NLO}
\end{eqnarray}
We follow the philosophy of Ref. \cite{Dawson:1998py}  and approximate the form factors in the
virtual corrections by the exact $m_t$ dependent quantities and include the full mass dependence
in ${\hat\sigma}_{LO}^{m_t}$ in Eq. \ref{eq:NLO}.
The coefficient, $C^{m_t}$,  for the virtual corrections is
\begin{eqnarray}
C^{m_t} & = & \pi^2 + {11\over 2} + \frac{33-2n_{lf}}{6} \log \frac{\mu_R^2}{s} 
\nonumber \\
&&+{8s\over 9 }\cos^2\theta Real \biggl(
{\int_{-{s\over 4}(\beta+1)^2}^{-{s\over 4}(\beta-1)^2} dt 
\biggl\{ F_1(s,t,u,m_t^2)-{p_T^2\over 2 t u}(s-2m_h^2) F_2(s,t,u,m_t^2)\biggr\}
\over
{\int_{-{s\over 4}(\beta+1)^2}^{-{s\over 4}(\beta-1)^2} dt 
\biggl\{
 \mid F_1(s,t,u,m_t^2)\mid^2 +
 \mid F_2(s,t,u,m_t^2)\mid^2  \biggr\}
 }\biggr)}
\label{eq:virtdef}
\end{eqnarray}
and 
\begin{eqnarray}
\beta&\equiv&\sqrt{1-{4m_h^2\over s}}\, .
\end{eqnarray}
 $P_{gg}(z)$ and $ P_{gq}(z)$ are the 
 DGLAP splitting functions,
\begin{eqnarray}
P_{gg}(z) &=& 6\left\{ \left(\frac{1}{1-z}\right)_+
+\frac{1}{z}-2+z(1-z) \right\} + \frac{33-2n_{lf}}{6}\delta(1-z), 
\nonumber \\
P_{gq}(z) &=& \frac{4}{3} \frac{1+(1-z)^2}{z},
\end{eqnarray}
where $n_{lf}=5$.  The result in Eq.~\ref{eq:NLO} has only approximate finite $m_t$ dependence since it has been adapted from the NLO calculation in the $m_t\rightarrow\infty$ limit~\cite{Dawson:1998py} .

We define an $m_t$ dependent differential $K$ factor from Eqs. \ref{sigmlo} and \ref{eq:NLO},
\begin{equation}
K^{m_t}\equiv \frac{d\sigma^{m_t}_{NLO}}{dM_{hh}}\bigg{/}\frac{d\sigma^{m_t}_{LO}}{dM_{hh}}\, ,
\label{eq:Kfin}
\end{equation}
where $M_{hh}$ is the invariant mass of the final state double Higgs system.  
In the following section, we will also show the numerical effects 
on the $K$ factor of replacing the form factors and LO cross section
by their $m_t\rightarrow\infty$ limits, 
\begin{equation}
K^{m_t\rightarrow\infty}
\equiv \frac{d\sigma^{m_t\rightarrow\infty}_{NLO}}{dM_{hh}}\bigg{/} \frac{d\sigma^{m_t\rightarrow \infty}_{LO}}{dM_{hh}}\, ,
\label{eq:Kinf}
\end{equation}
where,
\begin{eqnarray}
C^{m_t\rightarrow\infty} & = & \pi^2 + {11\over 2} + \frac{33-2n_{lf}}{6} \log \frac{\mu_R^2}{s} 
+{2\over 3}\cos^2\theta {Real \biggl( c_\Delta(s)-\cos^2\theta\biggr)
\over
 \mid c_\Delta(s)-\cos^2\theta\mid^2 }
\label{eq:virtdefinf}
\end{eqnarray}
and
\begin{equation}
c_\Delta=
\biggl({\cos\theta \lambda_{111}v\over s-m_h^2+im_h\Gamma_h}
+{\sin\theta \lambda_{211}v\over s-M_H^2+iM_H\Gamma_H}\biggr)\, .
\end{equation}

\section{Results}
\label{sec:results}
Our results are computed using CT12NLO PDFs\cite{Owens:2012bv} with a central scale choice $\mu_R=\mu_F\equiv\mu=M_{hh}$ for the
renormalization and factorization scales, and with $m_t=173.34~GeV$ and $m_b=4.62~GeV$.  In 
the computation of $\Gamma(H\rightarrow hh)$, we use the ${\overline{MS}}$ NNLO running
mass for ${\overline{m_b}}(M_H)$ and we always assume that $m_h=125~GeV$ is the lightest
Higgs boson.  Finally, the production cross section is computed including only the top quark loops, which are the largest contribution.  Our numerical results in the SM are checked using the 
program HPAIR\cite{Dawson:1998py}.  The singlet model results from our private code were checked by incorporating the resonance from the singlet model in HPAIR and comparing the two results.

\subsection{SM Results}
\begin{figure}
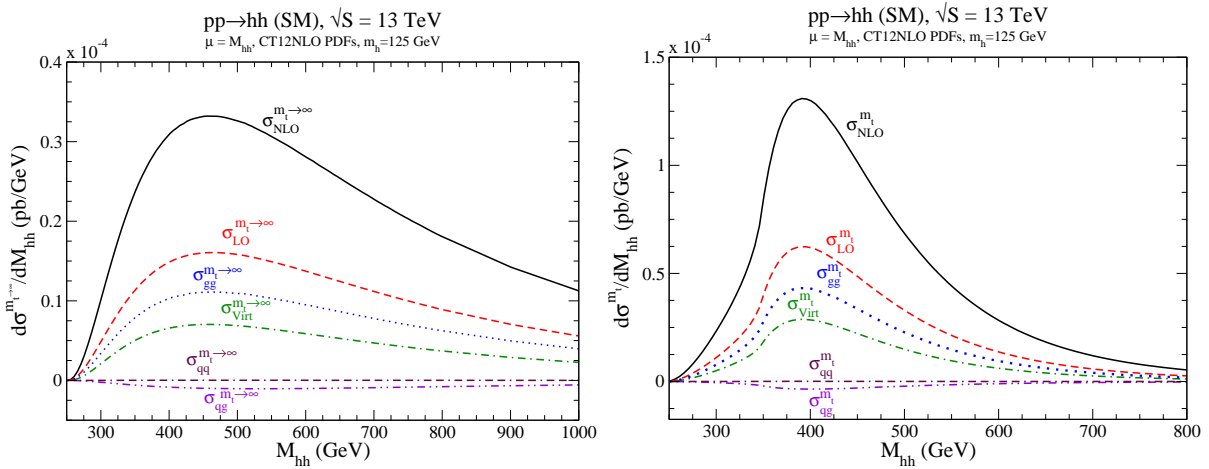

\begin{centering}
\includegraphics[width=0.48\textwidth,clip]{dsigdmhh_sm_mtinf.eps} 
\includegraphics[width=0.48\textwidth,clip]{dsigdmhh_sm.eps}
\par\end{centering}
\caption{Contributions to the SM rate for
 $pp\rightarrow h h$ at $\sqrt{S}=13~TeV$
in the $m_t\rightarrow \infty$ limit (LHS) and using the approximated $m_t$ dependence of Eqs. \ref{eq:NLO},\ref{eq:virtdef} (RHS).}
\label{fig:SM-diff}
\end{figure}

\begin{figure}
\begin{centering}
\includegraphics[width=0.45\textwidth,clip]{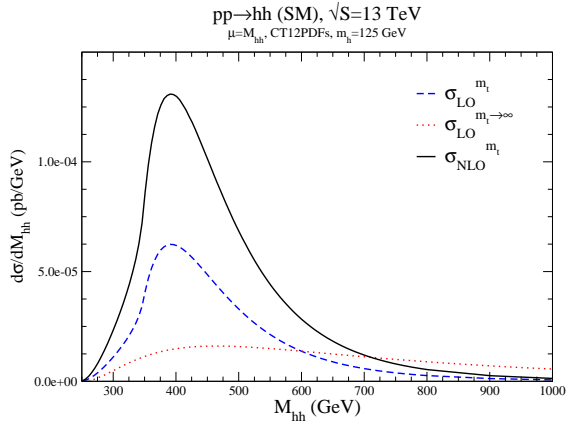}
\par\end{centering}
\caption{SM differential cross section for
 $pp\rightarrow h h$ at $\sqrt{S}=13~TeV$.   The NLO curve labelled $\sigma^{m_t}$ is obtained from Eqs. \ref{eq:NLO},\ref{eq:virtdef}. }
\label{fig:results_sm}
\end{figure}

The LO rate for $gg\rightarrow hh$ in the SM is well known, as are the NLO  and NNLO rates in the $m_t\rightarrow\infty$ limit. 
Ref. \cite{deFlorian:2015moa} finds the NNLO matched to NNLL rate of $36.8~fb$ for $pp\rightarrow hh$ at $\sqrt{S}=13 ~TeV$, $\mu=M_{hh}$, using
MSTW2008 PDFs.
The contributions to the differential SM NLO cross section are shown in Fig. \ref{fig:SM-diff}
in the $m_t\rightarrow \infty$ limit (LHS) and in the $m_t$ dependent
approximation of Eqs. \ref{eq:NLO} and \ref{eq:virtdef} (RHS).   The normalization and shapes of the $2$ approximations
are quite different, but the $K$ factors computed from the $2$ approximations are almost identical.
The contributions from real gluon emission, $\sigma_{gg}$,
and from the one-loop virtual diagrams, $\sigma_{\rm virt}$, are of similar sizes, while the contributions 
from quark initial states are highly suppressed.  
  In Fig.~\ref{fig:results_sm}, we show the NLO result with approximate $m_t$ dependence as defined in Eqs.~\ref{eq:NLO},~\ref{eq:virtdef}, and LO results for $m_t\rightarrow\infty$ and including the $m_t$ dependence
  exactly.
 The lowest order result in the $m_t\rightarrow\infty$ limit overshoots
 the exact lowest order result at high $M_{hh}$ and fails to reproduce the peak structure, as
 is well known.
    Including
 the NLO corrections significantly increases the rate.  (Calculating 
  $K^{m_t\rightarrow\infty}$ from the LHS of Fig. \ref{fig:SM-diff} and weighting by the exact $m_t$
 dependent LO result gives a curve which is almost indistinguishable from the NLO curve of Fig. \ref{fig:results_sm}.)
  
\begin{figure}
\begin{centering}
\includegraphics[width=0.45\textwidth,clip]{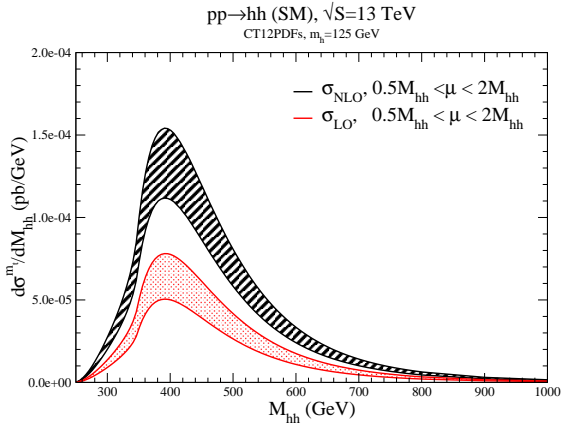}
\par\end{centering}
\caption{Scale dependence of the SM differential cross section for
 $pp\rightarrow h h$ at $\sqrt{S}=13~TeV$.  The NLO curves are obtained 
 using Eqs. \ref{eq:NLO},\ref{eq:virtdef}.}
\label{fig:scale_sm}
\end{figure}

\begin{figure}
\begin{centering}
\includegraphics[width=0.45\textwidth,clip]{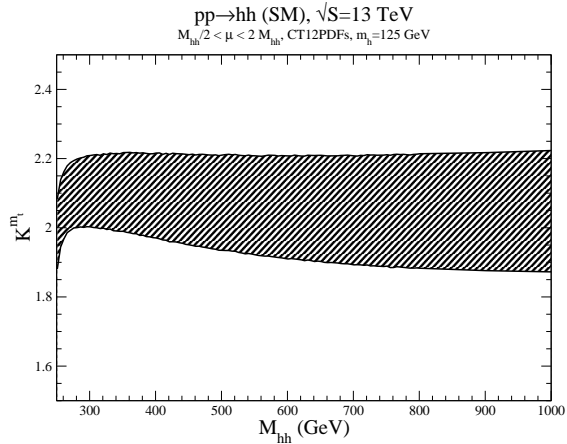}
\par\end{centering}
\caption{Scale variation of the SM $K^{m_t}$ factor, Eq.~\ref{eq:Kfin}, for
 $pp\rightarrow h h$ at $\sqrt{S}=13~TeV$
 when the scale is varied ${M_{hh}\over 2} < \mu<2M_{hh}$. }
 \label{fig:SM-kfac}
\end{figure}

 We show the 
 renormalization/factorization scale variation of the  SM LO and NLO
 rates in Fig. \ref{fig:scale_sm} when $M_{hh}/2<\mu<2 M_{hh}$.  In this figure, the LO rate includes all top mass dependence and the NLO rates
 are calculated using Eq. \ref{eq:NLO}. The fractional scale dependence is significantly reduced at NLO.
 The scale variation of the differential SM $K^{m_t}$ factor defined in Eq. \ref{eq:Kfin} is shown in Fig. 
 \ref{fig:SM-kfac}.  At $M_{hh}=400~GeV$, the NLO scale uncertainty is $\sim 11 \%$, while at 
 $M_{hh}=800~GeV$
 it is  $\sim 15\%$.   In the SM, the differential $K$ factor is only slightly dependent on $M_{hh}$ and
 can be accurately approximated by a constant.

\subsection{Singlet Model Results}

\begin{figure}
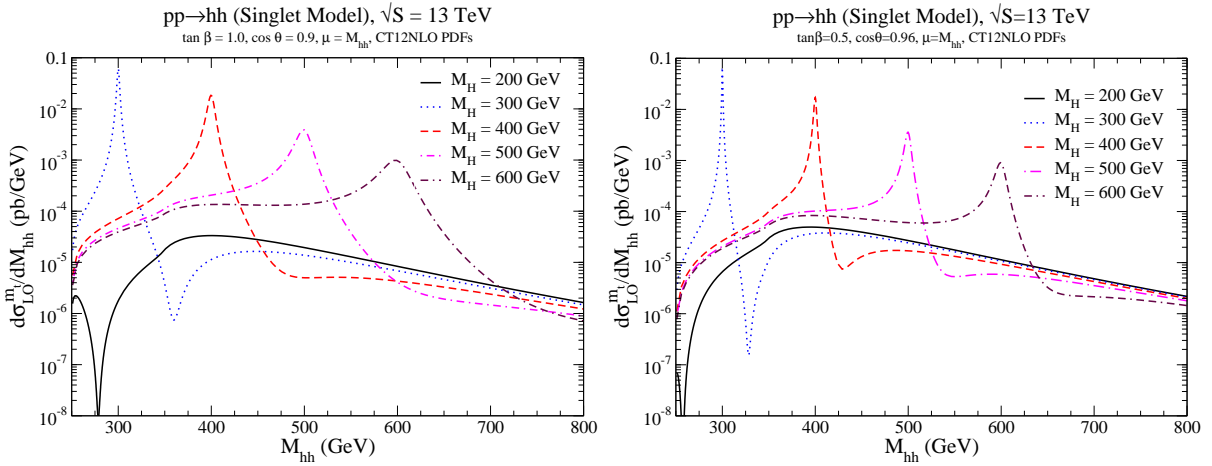

\begin{centering}
\includegraphics[width=0.48\textwidth]{dsig_singlet_masses_tb1_new.eps}
\includegraphics[width=0.48\textwidth]{dsig_singlet_masses_tb5.eps}
\par\end{centering}
\caption{Exact LO rates for
 $pp\rightarrow h h$ at $\sqrt{S}=13~TeV$
for fixed singlet mixing parameters, $\cos\theta=0.9$ and
$\tan\beta=1.0$ (LHS) and $\cos\theta=0.96$ and
$\tan\beta=0.5$ (RHS).}
\label{fig:singlet_tb1}
\end{figure}

\begin{figure}
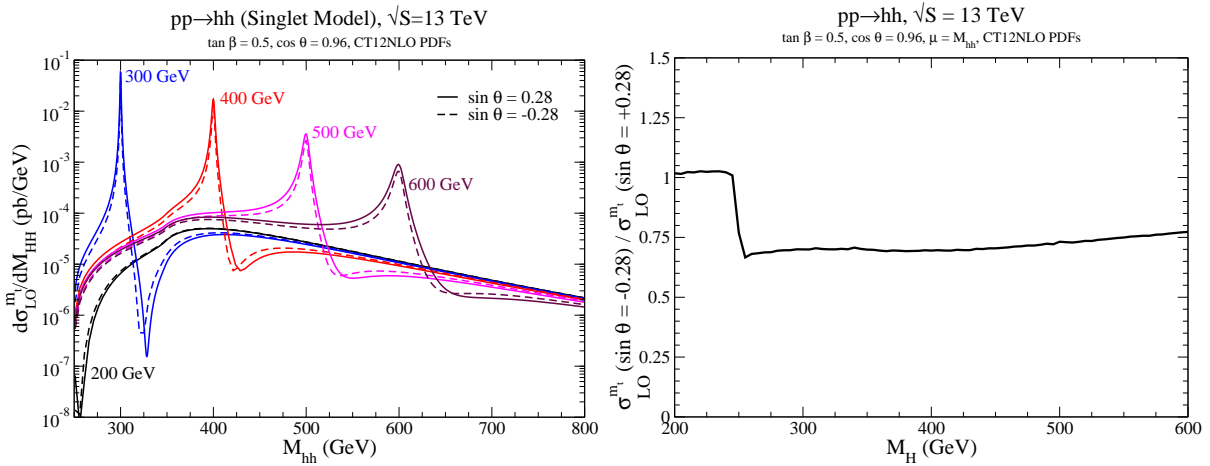

\begin{centering}
\includegraphics[width=0.48\textwidth]{dsig_singlet_masses_tb5_sign.eps}
\includegraphics[width=0.48\textwidth,clip]{sig_singlet_masses_tb05_signrat.eps}
\par\end{centering}
\caption{Exact LO rates for
 $pp\rightarrow h h$ at $\sqrt{S}=13~TeV$
for fixed singlet mixing parameters $\cos\theta=0.96$ and
$\tan\beta=0.5$.  (LHS) The solid (dashed) curves correspond to choosing a positive (negative) sign for $\sin\theta$.  (RHS) The ratio of the total LO cross sections with negative and positive sign for $\sin\theta$.}
\label{fig:sign}
\end{figure}

We begin by showing some lowest order  results.  The LO rate as a function
of $M_H$ is shown in Figs. \ref{fig:singlet_tb1} and \ref{fig:sign}. For the
smaller values of $\tan\beta$ and $\theta$, the resonances become narrower, while for heavier $M_H$ the height of 
the resonance peak and the dip above the peak due to interference effects become smaller.   The strength of the destructive interference is particularly strong for $M_H=200$~GeV. Interference effects will be more thoroughly discussed in the next section, but we give an outline here. As in the SM, the box diagram dominates the $h$-resonance. Hence, the major contributions to interference are between the $h$- and $H$-resonances and the box diagram.  The $h$-resonance and box diagrams (SM like contributions) have destructive interference, and the $H$-resonance and box diagrams have constructive (destructive) interference for $M_{hh}<M_H$ ($M_{hh}>M_H$).  In the SM, the $h$-resonance and box diagrams have exact destructive interference at the double-Higgs threshold.  In the singlet model the cancellation is not exact anymore due to changes in the tri-linear coupling and different mixing angle suppressions of the two diagrams, but the SM like contributions still have the strongest destructive interference at threshold.  For $M_H<2m_h$, both $h$- and $H$-resonance diagrams have strong destructive interference with the box diagram near $M_{hh}\sim2m_h$.  Hence, the overall destructive interference dip is strongest for $M_H=200$~GeV.

In picking a parameter point, we have a choice
as to whether to choose a positive or negative sign for $\sin\theta$.  The comparison of these two choices is shown in Fig.~\ref{fig:sign}, 
with the LHS showing the differential cross sections and the RHS the ratio of total cross sections. As shown in the LHS of Fig.~\ref{fig:sign}, the choice of sign makes little
difference in the shape of the distributions.  In particular, the interference effects remain essentially unchanged.  
  This can be understood by analyzing the triple couplings $\lambda_{111}$ (Eq.~\ref{lam111}) and $\lambda_{211}$ (Eq.~\ref{lam211}), and $F_1$ (Eq. \ref{formdef}).  The dependence of the cross section on 
the sign of $\sin\theta$ always appears with an associated factor of  $\tan\beta$ and is suppressed
 compared to the $\cos\theta$ terms in the triple couplings.  However, there can still be a significant change in the total rate, as shown on the RHS of Fig.~\ref{fig:sign}.  For $M_H>2m_h$, the cross section for negative $\sin\theta$ is $\sim70-80\%$ that of the cross section for positive $\sin\theta$.  For $M_H<2m_h$, the two cross sections are nearly the same.  This can be understood, and is shown later, by noting that the $H$ resonance makes a subleading contribution for $M_H<2m_h$~GeV and the SM like contributions only depend on the sign of $\sin\theta$ in a highly suppressed $\sin^3\theta$ term in $\lambda_{111}$.  Throughout the rest of the paper we will choose a positive sign for $\sin\theta$.

\begin{figure}
\begin{centering}
\includegraphics[width=0.48\textwidth]{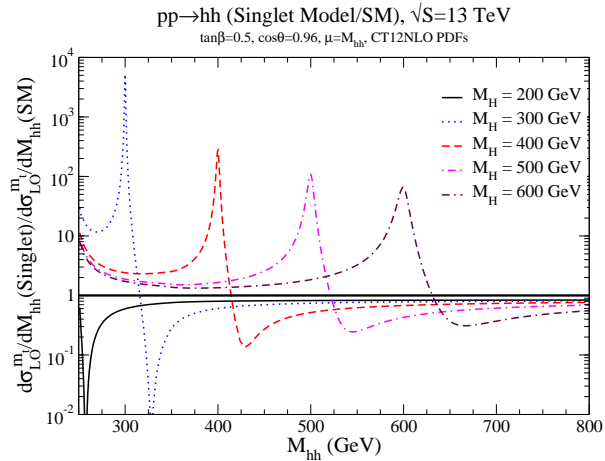}
\par\end{centering}
\caption{Exact LO rates for
 $pp\rightarrow h h$ normalized to the exact LO SM rate at $\sqrt{S}=13~TeV$
for fixed singlet mixing parameters, $\cos\theta=0.96$ and
$\tan\beta=0.5$.} 
\label{fig:rats}
\end{figure}

 In Fig.~\ref{fig:rats},
we show the ratio of the singlet model rate normalized to the SM rate.  It is
clear that near the resonances large enhancements in the rates are possible and the singlet model should be clearly
distinguishable from the SM.

\subsection{Interference effects}

The presence of the second scalar leads to interesting interference effects with the SM-like contributions.  
  The real parts of the propagators in $F^{tri}_1$ (Eq.~\ref{formdef}) interfere destructively for $m_h < M_{hh}< M_H$
and constructively for $M_{hh}> M_H$, as is typical for the interference of two 
resonances\footnote{This same interference effect is seen in the process $gg\rightarrow ZZ$ in the
singlet model\cite{Kauer:2015hia,Maina:2015ela} and in Drell-Yan production below the $Z$ peak.}.   However, in the SM the box and triangle diagrams
 destructively interfere, with the box diagram dominating at large $M_{hh}$\cite{Chen:2014xra}.
   Hence, although the propagators of the two resonances 
destructively interfere, the $H$-propagator constructively interferes with the box diagram for $M_{hh} < M_H$, and destructively interferes for $M_{hh}>M_H$.

\begin{figure}
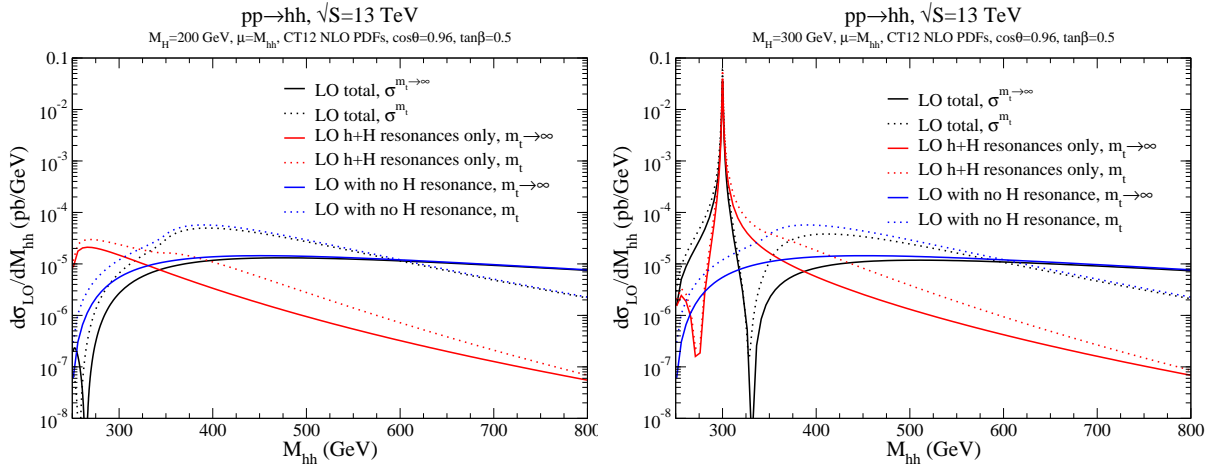

\begin{centering}
\includegraphics[width=0.48\textwidth]{res_200_tb5.eps}
\includegraphics[width=0.48\textwidth]{res_300_tb5.eps}
\par\end{centering}
\caption{LO results for $pp\rightarrow hh$ at $\sqrt{S}=13~TeV$ for
$M_H=200~GeV$ (LHS) and $M_H=300~GeV$ (RHS),
$\cos\theta=0.96$ and
$\tan\beta=0.5$.  See text for description of individual curves.}
\label{fig:int200}
\end{figure}

\begin{figure}
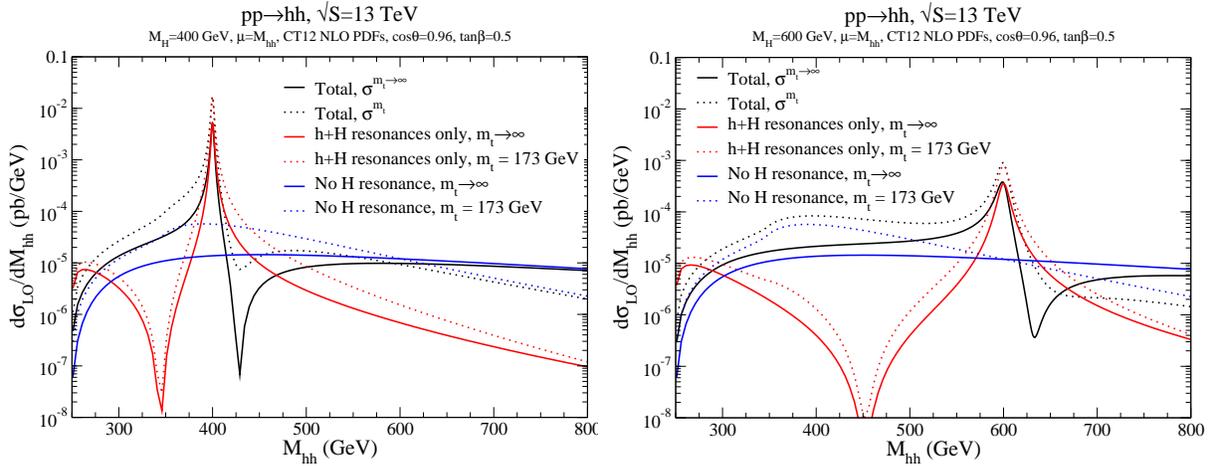

\begin{centering}
\includegraphics[width=0.48\textwidth]{res_400_tb5.eps}
\includegraphics[width=0.48\textwidth]{res_600_tb5.eps}
\par\end{centering}
\caption{LO results for $pp\rightarrow hh$ at $\sqrt{S}=13~TeV$ for
$M_H=400~GeV$ (LHS), $M_H=600~GeV$ (RHS), with
$\cos\theta=0.96$ and
$\tan\beta=0.5$.  See text for description of individual curves.}
\label{fig:int400}
\end{figure}

Leading order differential cross sections with individual contributions are shown separately  in Figs. \ref{fig:int200} and \ref{fig:int400}. 
The curves labelled ``$h+H$ resonances only" include the contributions of both $s$-channel $h$ and $H$ and their
interference, but not the effect of the box diagrams.  The destructive interference between the two propagators for
$M_{hh}<M_H$ is clear.  The curves labelled ``no $H$-resonance" have the $H$ resonance 
contribution removed; that is, only the SM-like contributions are included.  As described above, by comparing the curves labelled 
``no H-resonance" with the total distribution, we see that there is constructive interference between the $H$ and SM-like diagrams for 
$M_{hh}<M_H$ and destructive interference for $M_{hh}>M_H$.  
It is apparent that the $m_t\rightarrow\infty$ limit fails to reproduce the correct interference structure near and slightly above the peak
and overshoots the rate at high $M_{hh}$.  
The location of the interference dip just above the resonance is slightly shifted to larger $M_{hh}$ in the $m_t\rightarrow\infty$ limit.
This motivates weighting the NLO rate (which is only known in the $m_t\rightarrow\infty$ limit),
by the exact LO rate.

\begin{figure}
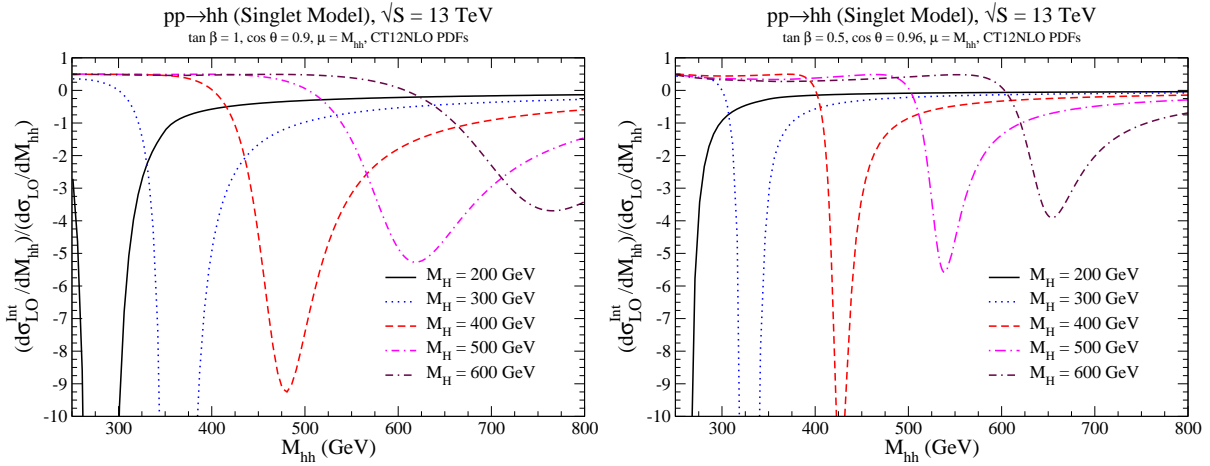

\begin{centering}
\includegraphics[width=0.48\textwidth]{Interference.eps}
\includegraphics[width=0.48\textwidth]{Interference_tb05.eps}
\par\end{centering}
\caption{Ratio of the interference between the $H$-resonance and the SM-like contributions,
$h$-resonance+box, and the full invariant mass distribution with $\tan\beta=1.0$ and $\cos\theta=0.9$ (LHS), and $\tan\beta=0.5$ and $\cos\theta=0.96$ (RHS).
The curves use the exact LO matrix elements.}
\label{fig:inter}
\end{figure}

We show the ratio of the interference between 
the $H$ resonance and SM-like diagrams and the full invariant mass distribution in Fig.~\ref{fig:inter}.  Exact $m_t$ dependence has been kept.  The interference contribution is
\begin{eqnarray}
\frac{d\sigma^{Int}_{LO}}{dM_{hh}}=\frac{d\sigma_{LO}}{dM_{hh}}-\left(\frac{d\sigma^H_{LO}}{dM_{hh}}+\frac{d\sigma^{h+Box}_{LO}}{dM_{hh}}\right),
\label{eq:int}
\end{eqnarray}  
where $\sigma^H$ contains only the contribution from the $H$-resonance, and $\sigma^{h+Box}$ contains the $h$-resonance and box contributions and their interference.  An interesting feature of Fig.~\ref{fig:inter} is that for $M_{hh}\ll M_H$, the interference 
contribution is independent of $M_H$ for fixed $\theta$ and $\tan\beta$.  This somewhat surprising effect can be understood 
by taking $F_1$ (Eq.~\ref{formdef}) in the limit $m_h^2,s\ll M_H^2$: 
\begin{eqnarray}
F_1&\rightarrow &-s \biggl(\frac{\cos\theta \lambda_{111}v}{s-m_h^2+im_h\Gamma_h}
+\frac{\sin\theta\sin2\theta}{2}(\cos\theta+\sin\theta\tan\beta)\biggr)F_{\triangle}(s,m_t^2)\nonumber \\
&&
+s\cos^2\theta F_{\square}(s,t,u,m_t^2).
\end{eqnarray}
As can be clearly seen, in this limit, the double Higgs rate does not explicitly depend on the heavy scalar mass.

\begin{figure}
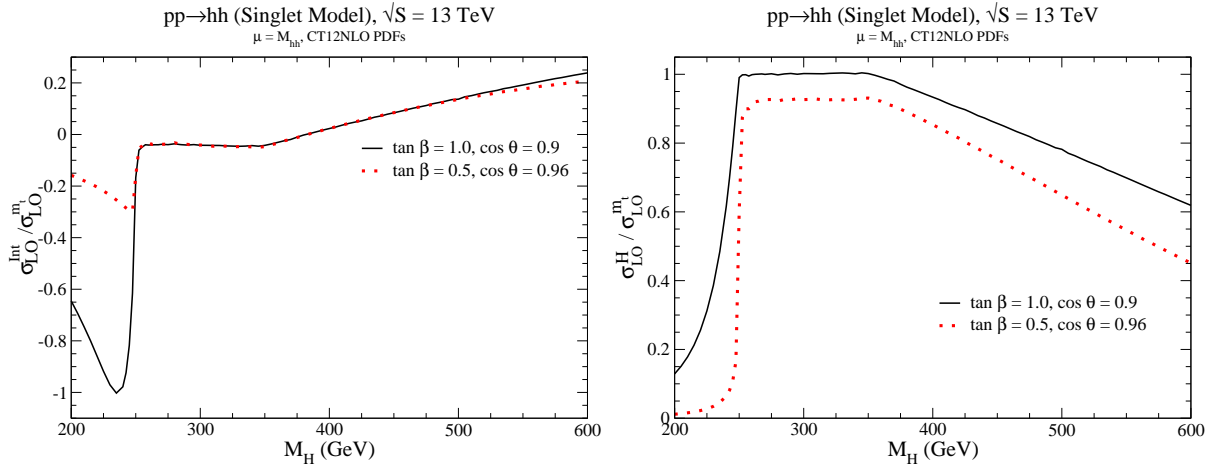

\begin{centering}
\includegraphics[width=0.48\textwidth,clip]{Interference_tot.eps}
\includegraphics[width=0.48\textwidth,clip]{Res_frac_tot.eps}
\par\end{centering}
\caption{Ratio of the interference between the $H$-resonance and the SM-like contributions,
$h$-resonance+box, and the total cross section as a function of $M_H$ (LHS).  Ratio of the H-resonance contribution only and the total rate (RHS).  Both $\tan\beta=1.0,\cos\theta=0.9$ (solid black) and $\tan\beta=0.5,\cos\theta=0.96$ (dotted red) are shown.
The curves use the exact LO matrix elements.}
\label{fig:intertot}
\end{figure}

The ratio of the interference between the $H$-resonance and SM-like contributions defined in Eq.~\ref{eq:int} and the total cross section are shown in the LHS of Fig.~\ref{fig:intertot}.  We also show the ratio of the $H$-resonance contribution only and the total cross section in the RHS of Fig.~\ref{fig:intertot}.  The curves are shown for the two parameter points $\tan\beta=1$, $\cos\theta=0.9$ (solid black) and $\tan\beta=0.5$, $\cos\theta=0.96$ (dotted red).  At amplitude level, the dominant (box) contribution to the SM-like pieces is proportional to $\cos^2\theta$ and makes a similar contribution for both parameter points.  However, below $2m_h$ the $H$-resonance amplitude is proportional to $\sin^2\theta$ and sensitive to relatively small changes in $\cos\theta$. 
This explains why for $M_H<2m_h$ the interference and $H$-resonance contributions are larger for $\cos\theta=0.9$ than for $\cos\theta=0.96$. For $M_H>2m_h$ and using the narrow-width-approximation, the $H$-resonance amplitude is proportional to $\sin\theta$ and is still larger for $\cos\theta=0.9$ than for $\cos\theta=0.96$. Once the resonance production of $hh$ turns on, $M_H\sim 2m_h$, the $H$-resonance contribution dominates, as seen in the RHS of Fig.~\ref{fig:intertot}.  As $M_H$ increases, the $H$-propagator suppresses the $H$-resonance contribution.  However, as $M_H$ approaches $2m_t$, as is well-known in single Higgs production, the production rate through a top quark triangle increases.  For $2m_h\lesssim M_H\lesssim 2m_t$ these two effects cancel each other and the contribution from the $H$-resonance is relatively constant.   As   $M_H$ increases above $\sim2m_t$, the suppression from the $H$-propagator is the dominant effect.  Hence, the fractional contribution from the $H$-resonance only decreases and the fractional contribution from interference increases.  These two effects are correlated because the SM-like contribution by itself is independent of $M_H$.  It should be noted that the absolute contribution from the interference is nearly independent of $M_H$ for $M_H\gtrsim500$~GeV.  This can be understood from Eq.~\ref{eq:int}.  Since for increasing $M_H$ there is a large contribution to the cross section from the $M_H\gg M_{hh}$ region, the total contribution to the interference is largely independent of $M_H$.

\subsection{NLO Effects}

\begin{figure}
\begin{centering}
\includegraphics[width=0.48\textwidth]{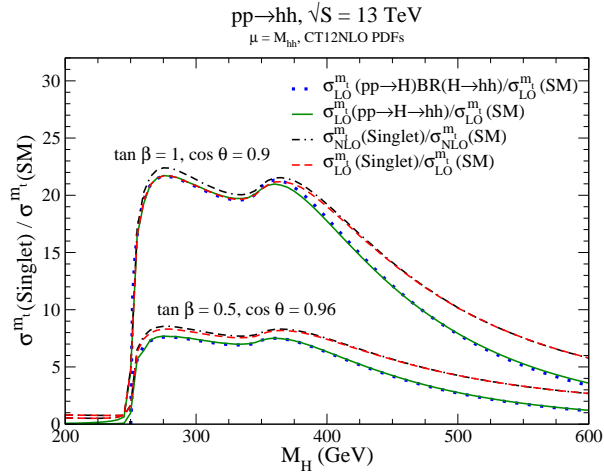}
\par\end{centering}
\caption{Ratio of NLO (dash-dot black) and LO (dashed red) rates to the respective SM rates.  Also, the ratio of the LO H-resonance only contribution calculated using the
 Breit-Wigner resonance (solid green) and in the narrow-width approximation (blue dotted) to the  LO SM rate.}
\label{sigtotrat}
\end{figure}
In Fig. \ref{sigtotrat}, we show the enhancement of the total cross section in the singlet model, relative to the SM rate.  For $\tan\beta=5$ and $\cos\theta=0.96$, the maximum enhancement is of ${\cal {O}}(8)$ for $M_H\lsim 500~GeV$ and decreases rapidly to  ${\cal{O}}(1)$ for larger $M_H$.  For larger mixing, $\tan\beta=1$ and $\cos\theta=0.9$, 
enhancements of the SM rate up to a factor of $\sim 22$ are possible.  We see that $\sigma^{m_t}/\sigma_{SM}$ is not very different for LO and NLO total rates.  The
contribution of the $H$ resonance in the narrow width approximation is accurate for $M_H\lsim 400~GeV$, but underestimates the enhancement for larger $M_H$. 

\begin{figure}
\begin{centering}
\includegraphics[width=0.48\textwidth]{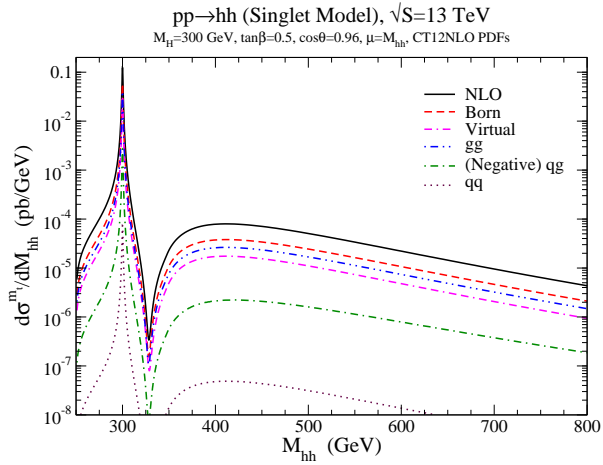}
\par\end{centering}
\caption{Total and individual contributions to the NLO cross sections defined in Eqs. \ref{eq:NLO},\ref{eq:virtdef}.  The model parameters were chosen to be  $\tan\beta = 0.5$, $\cos\theta=0.96$, and $M_H=300~GeV$.}
\label{fig:nloconts}
\end{figure}

We now present our numerical results for the double Higgs invariant mass distributions at NLO. Fig.~\ref{fig:nloconts} shows the 
individual contributions (Eq.~\ref{eq:NLO}) to the invariant mass distributions using the approximation of Eqs. \ref{eq:NLO},\ref{eq:virtdef}.
 It is important to remember that the full $m_t$ dependent NLO rate is not known.  We plot the absolute value of 
the $qg$ contribution, since it is negative.  The leading corrections are from the $gg$ and virtual contributions, while the $qg$ and $q{\overline{q}}$
contributions are subleading.  

\begin{figure}
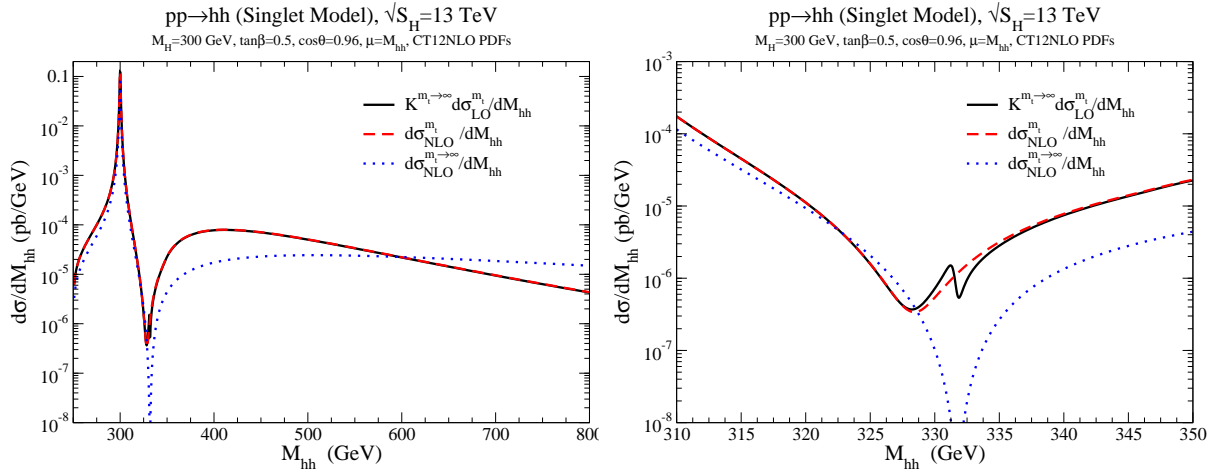

\begin{centering}
\includegraphics[width=0.48\textwidth]{dsigdmhh_300_NLO_comp_tb05.eps}
\includegraphics[width=0.48\textwidth]{dsigdmhh_300_NLO_comp_tb05_zoom.eps}
\par\end{centering}
\caption{NLO cross sections for $M_H=300$ GeV and with different approximations for the top mass
dependence as described in the text.  The mixing parameters were set to $\tan\beta=0.5$ and $\cos\theta=0.96$.}
\label{fig:nlocom}
\end{figure}

It is interesting to compare the effect of the approximations to the top mass dependence at NLO.  In Fig.
\ref{fig:nlocom}, we compare the NLO rate for $M_H=300~GeV$  computed using the approximation of Eqs. \ref{eq:NLO},\ref{eq:virtdef} (dashed red curve) with that obtained by computing
$K^{m_t\rightarrow\infty}$ (Eq.~\ref{eq:Kinf}) and weighting by the exact $m_t$ dependent LO cross section (solid black).  
The curves overlap almost exactly.  Since most contributions to the NLO rate (Eq.~\ref{eq:NLO}) are proportional to the LO rate, the approximate $m_t$ dependence is mostly captured by weighting the exact LO rate with $K^{m_t\rightarrow\infty}$.  The only complication is a piece of the virtual contribution (Eq.~\ref{eq:virtdef}) that is not proportional to the LO rate.  However, this piece turns out to make a small contribution.  

We then compare with an NLO rate computed in the $m_t\rightarrow \infty$ limit (dotted blue in Fig.~\ref{fig:nlocom}), i.e. this result is not reweighted
by the exact $m_t$ dependent LO result. The $m_t\rightarrow\infty$ limit shifts the location of the interference dip to slightly higher $M_{hh}$.
This effect is also apparent in the comparison of the exact $m_t$ dependent and $m_t\rightarrow\infty$ LO curves of Fig. \ref{fig:int200}.  A blow-up of the interference region is
shown on the RHS of Fig. \ref{fig:nlocom}  and makes this effect obvious.  

On the RHS of Fig. \ref{fig:nlocom}, we can also see that the curve obtained by weighting the exact LO rate by $K^{m_t\rightarrow\infty}$ differs from the curve calculated using Eqs.~\ref{eq:NLO},\ref{eq:virtdef} at the interference dip of the $m_t\rightarrow\infty$ curve.  The interference dip is where the LO cross section is a minimum.  Hence, the piece of the virtual contribution (Eqs. \ref{eq:virtdef},\ref{eq:virtdefinf}) not proportional the LO cross section makes a relatively large contribution in this region. Since the interference dip is deeper in the $m_t\rightarrow\infty$ limit (see Fig.~\ref{fig:int200}), this effect is more pronounced in the $m_t\rightarrow\infty$ case.  As a consequence, at the interference dip, the $m_t\rightarrow\infty$ NLO rate is not approximately proportional to the $m_t\rightarrow\infty$ LO rate.  Therefore, weighting the exact LO rate by $K^{m_t\rightarrow\infty}$  does not reproduce the curves computed using Eqs.~\ref{eq:NLO},\ref{eq:virtdef} precisely where the $m_t\rightarrow\infty$ rate has the strongest destructive interference.

\begin{figure}
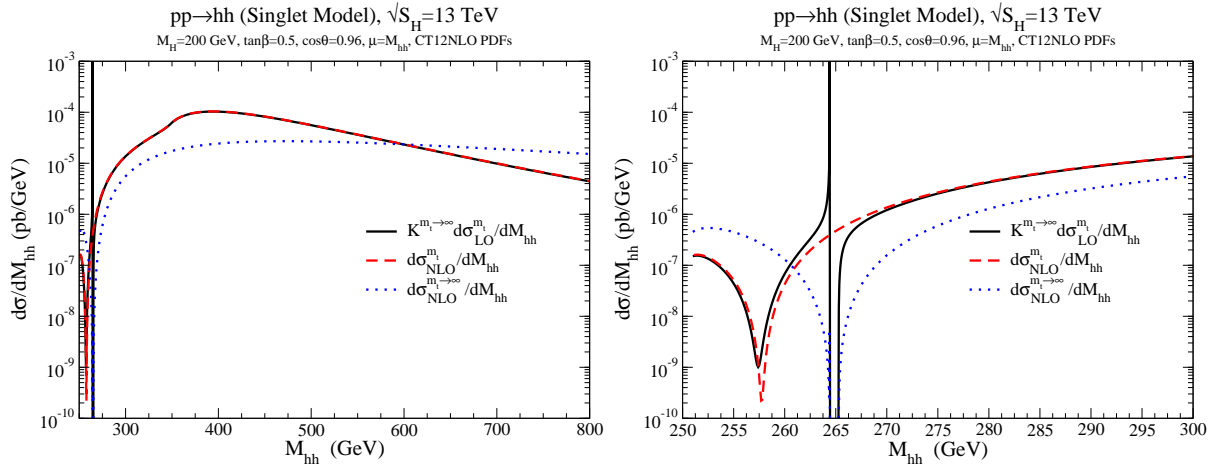

\begin{centering}
\includegraphics[width=0.48\textwidth]{dsigdmhh_200_NLO_comp_tb05.eps}
\includegraphics[width=0.48\textwidth]{dsigdmhh_200_NLO_comp_tb05_zoom.eps}
\par\end{centering}
\caption{NLO cross sections for $M_H=200$ GeV and with different approximations for the top mass
dependence as described in the text. The mixing parameters were set to $\tan\beta=0.5$ and $\cos\theta=0.96$.}
\label{fig:nlocom_200}
\end{figure}
   
\begin{figure}
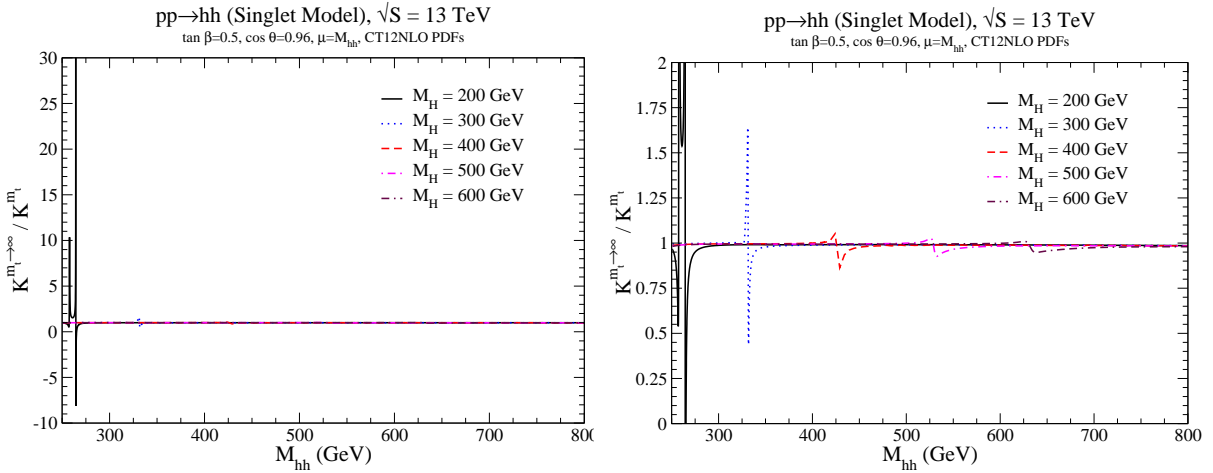

\begin{centering}
\includegraphics[width=0.48\textwidth]{KFactor_ratio_tb05.eps}
\includegraphics[width=0.48\textwidth]{KFactor_ratio_tb05_zoom.eps}
\par\end{centering}
\caption{Ratio of differential K-factors evaluated in the 
$m_t\rightarrow\infty$ limit to those calculated using the approximate $m_t$ dependence of Eq. \ref{eq:NLO}. The mixing parameters were set to $\tan\beta=0.5$ and $\cos\theta=0.96$.}
\label{fig:kfacrat}
\end{figure}

It is interesting to compare with the NLO rate for a heavy Higgs mass below the threshold for a double Higgs resonance, $M_H=200~GeV$.  These results are shown
in Fig. \ref{fig:nlocom_200}.  In the interference region, the effects are similar, but more pronounced, to those in the $M_H=300~GeV$ case.    In fact, for $M_H=200~GeV$, the two curves computed by Eqs.~\ref{eq:NLO},\ref{eq:virtdef} and by weighting the exact LO rate by $K^{m_t\rightarrow\infty}$ do not agree at the minimum of the $\sigma^{m_t}_{NLO}$ curve in addition to the minimum of the $K^{m_t\rightarrow\infty}\sigma^{m_t}_{LO}$ curve. This can be understood by noting that as $M_H$ increases, the interference dip of the LO cross section is more shallow (see Fig.~\ref{fig:singlet_tb1}).  As a consequence and discussed above, as $M_H$ increases the contribution to $\sigma_{virt}$ that is not proportional to the leading order rate decreases. Hence, the curves computed using Eqs.~\ref{eq:NLO},\ref{eq:virtdef} and weighting the exact LO rate with $K^{m_t\rightarrow\infty}$ will be in better agreement with increasing $M_H$.  In Fig.~\ref{fig:kfacrat} we show the ratio of the $K^{m_t\rightarrow\infty}$ and $K^{m_t}$ (Eq.~\ref{eq:Kfin}), which is the same as the ratio of the NLO rates calculated by weighting of the exact LO rate by $K^{m_t\rightarrow\infty}$ and using Eqs.~\ref{eq:NLO},\ref{eq:virtdef}.  As can be seen, as $M_H$ increases the two methods increasingly agree.

\begin{figure}
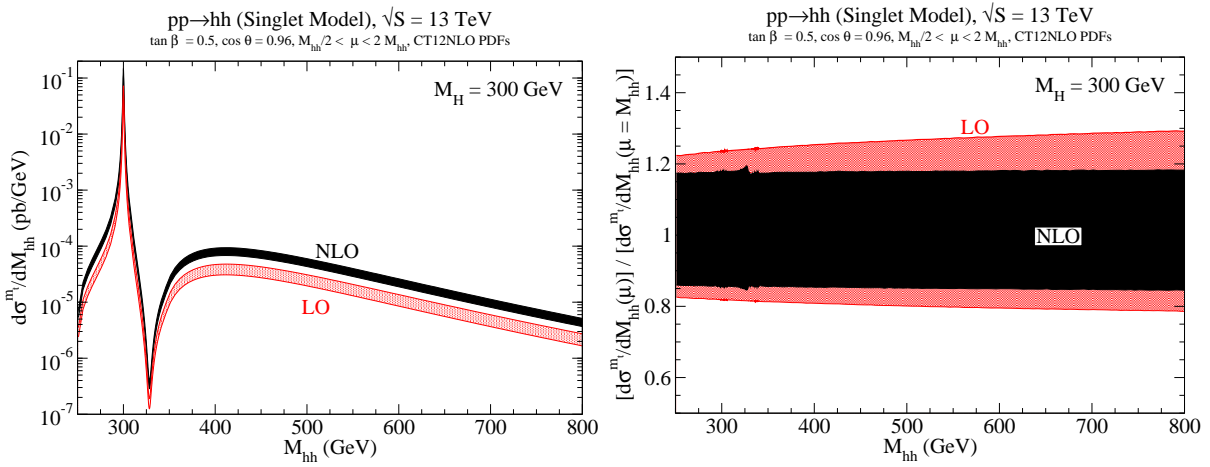

\begin{centering}
\includegraphics[width=0.48\textwidth]{dsig_scale.eps}
\includegraphics[width=0.48\textwidth]{dsig_scale_ratio.eps}
\par\end{centering}
\caption{Scale dependence of the invariant mass distributions for $M_H=300$~GeV for both the NLO and LO cross sections with $M_{hh}/2<\mu<2 M_{hh}$.  
The RHS shows the fractional scale dependence of the invariant mass distributions. These figures are computed using Eq. \ref{eq:NLO}.  The mixing parameters were set to $\tan\beta=0.5$ and $\cos\theta=0.96$.}
\label{fig:scale300}
\end{figure}

In Fig.~\ref{fig:scale300}, we show the scale dependence of the invariant mass distribution for a representative parameter point with $M_{hh}/2<\mu<2M_{hh}$.  The LO cross sections contains exact $m_t$ dependence and the NLO cross section is computed using Eqs.~\ref{eq:NLO},\ref{eq:virtdef}.  The NLO 
corrections decrease the scale dependence from $\sim^{+(20-30)\%}_{-20\%}$ to $\sim\pm15\%$.  Additionally, the NLO scale dependence 
is fairly flat throughout the distribution; in particular, it does not appreciably change in the resonance and strong destructive interference regions.

\begin{figure}
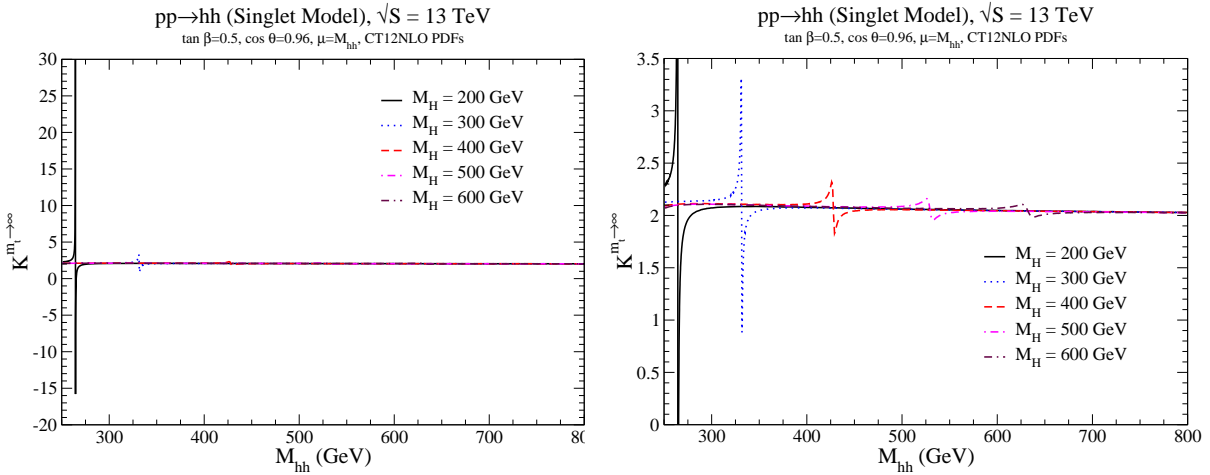

\begin{centering}
\includegraphics[width=0.48\textwidth]{KFactor_MtInf_tb05.eps}
\includegraphics[width=0.48\textwidth]{KFactor_MtInf_tb05_zoom.eps}
\par\end{centering}
\caption{$K^{m_t\rightarrow\infty}$ as defined in Eq.~\ref{eq:Kfin} at $\sqrt{S}=13~TeV$ for various $M_H$ and $\tan\beta=0.5$, $\cos\theta=0.96$}
\label{fig:kfac}
\end{figure}

In Fig.~\ref{fig:kfac}, we show the differential K-factor in the $m_t\rightarrow \infty$ limit, $K^{m_t\rightarrow\infty}$, as defined in Eq.~\ref{eq:Kinf}.  The K-factor is flat with a value of $2-2.2$, except 
for spikes that occur in the regions with the strongest destructive interference.  As shown in Fig.~\ref{fig:kfacrat}, the K-factor computed using Eq.~\ref{eq:Kfin} agrees with $K^{m_t\rightarrow\infty}$, except in the regions of strong destructive interference.

\subsection{Results at $100~TeV$}

\begin{figure}
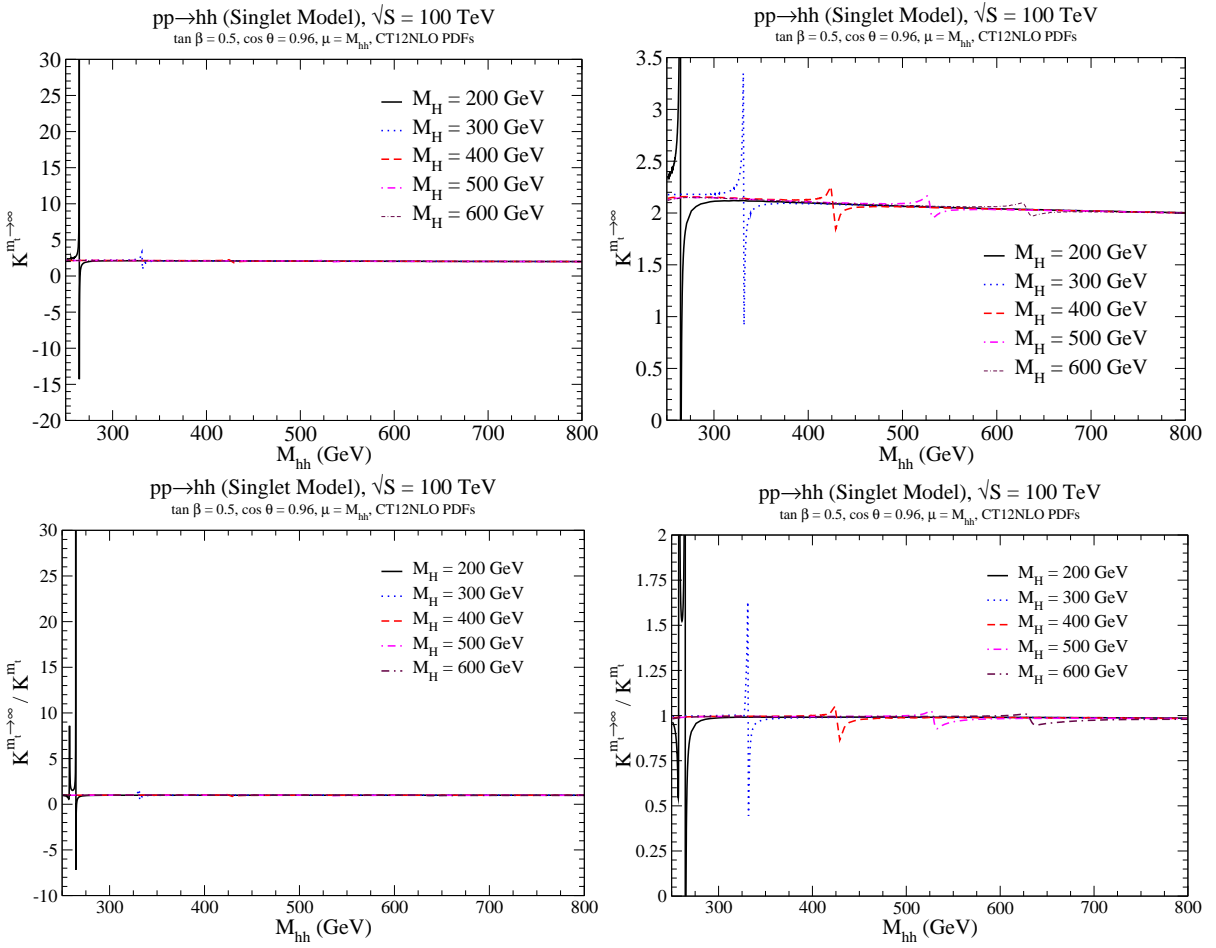

\begin{centering}
\includegraphics[width=0.48\textwidth,clip]{KFactor_MtInf_tb05_100TeV.eps}
\includegraphics[width=0.48\textwidth,clip]{KFactor_MtInf_tb05_100TeV_zoom.eps}\\
\includegraphics[width=0.48\textwidth,clip]{KFactor_ratio_tb05_100TeV.eps}
\includegraphics[width=0.48\textwidth,clip]{KFactor_ratio_tb05_100TeV_zoom.eps}
\par\end{centering}
\caption{(Top) $K^{m_t\rightarrow\infty}$ defined in Eq.~\ref{eq:Kinf}.  (Bottom) Ratio of $K^{m_t\rightarrow\infty}$ and $K^{m_t}$ defined in Eq.~\ref{eq:Kfin}. The mixing parameters were set to $\tan\beta=0.5$ and $\cos\theta=0.96$}
\label{fig:kfac100}
\end{figure}

Next we present our results for the NLO calculation of double Higgs production
 at $\sqrt{S}=100$~TeV.  In Fig.~\ref{fig:kfac100} we plot (top) $K^{m_t\rightarrow\infty}$, Eq.~\ref{eq:Kinf}, and (bottom) the ratio of the $K^{m_t\rightarrow\infty}$ and $K^{m_t}$, Eq.~\ref{eq:Kfin}.  The K-factors at $\sqrt{S}=100$ TeV are similar to those at $\sqrt{S}=13$~TeV.  Since the ratio of K-factors at 100 TeV is similar to those at 13 TeV, our comparison of the rates calculated by weighting the exact LO rate by $K^{m_t\rightarrow\infty}$ and Eqs.~\ref{eq:NLO},\ref{eq:virtdef} will translate from the 13 TeV to 100 TeV environment.

\begin{figure}
\begin{centering}
\includegraphics[width=0.45\textwidth]{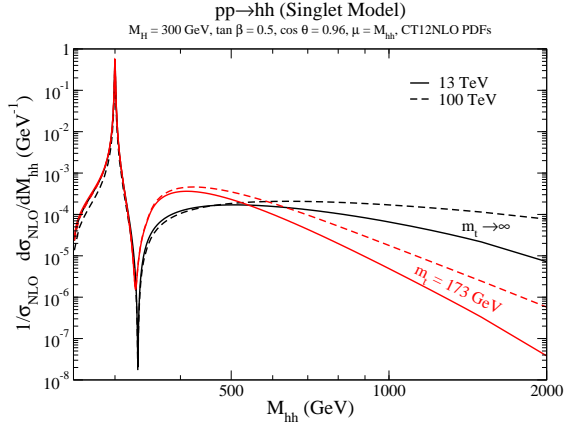}
\par\end{centering}
\caption{Normalized invariant mass distributions for $\sqrt{S}=13$ and $100$ TeV in both
the $m_t\rightarrow\infty$ limit and the approximate 
$m_t$ dependence of Eq.\ref{eq:NLO}.  The model parameters were set to $M_H=300~GeV$, $\tan\beta=0.5$ and $\cos\theta=0.96$.}
\label{fig:dsignorm}
\end{figure}

In Fig.~\ref{fig:dsignorm} we show the normalized invariant mass distributions at $\sqrt{S}=13$ and $100$ TeV with both $m_t\rightarrow \infty$ and the  approximate finite $m_t$
dependence of Eqs. \ref{eq:NLO},\ref{eq:virtdef}.  
As noted previously, the infinite top quark mass limit overestimates the tail of the distribution. 
Additionally, for the SM-like contributions, the $m_t\rightarrow\infty$ limit underestimates the cross section for $M_{hh}\lesssim550$~GeV (Fig.~\ref{fig:results_sm}).  Hence, after the strongest destructive interference, the SM-like contribution to the approximate finite $m_t$ rate grows more quickly than in the $m_t\rightarrow\infty$ case.  As a result, directly after the interference dip, the approximately finite $m_t$ distribution grows more quickly and obtains a higher value  than the $m_t\rightarrow\infty$ distribution.  
 Finally, at $\sqrt{S}=100$ TeV the tails of the distributions are enhanced relative to $13$~TeV.  This is because for a given invariant mass, the PDFs are evaluated at smaller $x$ at $100$ TeV than at $13$ TeV.  Hence, the enhancement of the gluon parton luminosity causes the tail of the distribution to be longer.

\section{Conclusions}
The production of Higgs pairs from gluon fusion is an important probe of the structure of the scalar potential.  In the SM, the QCD corrections are known in an approximation where
the LO rate is weighted by a $K$ factor computed in the $m_t\rightarrow \infty$ limit, increasing both the total rate and $d\sigma/dM_{hh}$ by a factor of around $2$.

We have presented results in the Higgs singlet model, where the tri-linear Higgs self coupling is modified from the SM value and significant resonant effects from the second scalar occur. The effects of the interference between the heavy scalar and SM-like contributions can be significant, altering invariant mass distributions for all $M_H$. For $M_H\gtrsim 450$~GeV, the interference effects can make a $\sim 10-20\%$ contribution to the total rate.  For $M_H\lesssim 2m_h$, the interference effects can suppress the total cross section up to $\sim 30\%$ for a viable parameter point.  Hence, in searches for heavy scalars, these effects should be included.

We compare an approximation for the NLO QCD corrections where the exact $m_t$ dependent LO cross section is weighted by a $K$ factor computed in the $m_t\rightarrow\infty$ limit, and alternatively where the exact $m_t$ dependent form factors are inserted into
the NLO contributions.  The approaches give similar results except in the regions with large destructive interference. 

In the singlet model, the total cross section is increased
by factors between $5-10$ above the SM rate for $\tan \beta=0.5$ and $\cos\theta=0.96$.  For larger mixing ($\tan\beta=1$ and $\cos\theta=0.9$), we find enhancements
from the SM rate  between $10-20$ for $M_H < 500 ~GeV$, and
the enhancement is very similar  at LO and NLO.  The resonant approximation to the total cross section underestimates the enhancement by about a 
factor of $2$ at large $M_H$.

The singlet model demonstrates a case where the kinematic distributions of the outgoing SM Higgs pair are significantly altered from the SM, and where the
higher order QCD corrections differ from those of the SM near the resonance peak.

\section*{Acknowledgements}
This work is supported by the U.S. Department of Energy under grant
No.~DE-AC02-98CH10886 and contract DE-AC02-76SF00515.  We thank
Chien-Yi Chen and Tania Robens for discussions.

\bibliographystyle{h-physrev}
\bibliography{hh}

\end{document}